\documentclass[twoside,twocolumn,english,aps,amsmath,amssymb,reprint,pra,superscriptaddress]{revtex4-1}
\usepackage[T1]{fontenc}
\usepackage[latin9]{inputenc}
\usepackage{color}
\usepackage{amsmath}
\usepackage{graphicx}

\makeatletter

\providecommand{\tabularnewline}{\\}

\usepackage{babel}

\usepackage{babel}

\makeatother

\usepackage{babel}
\begin{document}

\title{Memory-Assisted Quantum Key Distribution with a Single Nitrogen Vacancy
Center}

\author{Nicol$\acute{{\rm {o}}}$ Lo Piparo}

\affiliation{School of Electronic and Electrical Engineering, University of Leeds,
Leeds, UK}

\affiliation{National Institute of Informatics, 2-1-2 Hitotsubashi, Chiyoda, Tokyo
101-0003, Japan.}

\author{Mohsen Razavi}

\affiliation{School of Electronic and Electrical Engineering, University of Leeds,
Leeds, UK}

\author{William J. Munro}

\affiliation{National Institute of Informatics, 2-1-2 Hitotsubashi, Chiyoda, Tokyo
101-0003, Japan.}

\affiliation{NTT Basic Research Laboratories, NTT Corporation, 3-1 Morinosato-Wakamiya,
Atsugi, Kanagawa, 243-0198, Japan.}
\begin{abstract}
Memory-assisted measurement-device-independent quantum key distribution
(MA-MDI-QKD) is a promising scheme that aims to improve the rate-versus-distance
behavior of a QKD system by using the state-of-the-art devices. It
can be seen as a bridge between current QKD links to quantum repeater
based networks. While, similar to quantum repeaters, MA-MDI-QKD relies
on quantum memory (QM) units, the requirements for such QMs are less
demanding than that of probabilistic quantum repeaters. Here, we present
a variant of MA-MDI-QKD structure that relies on only a single physical
QM: a nitrogen-vacancy center embedded into a cavity where its electronic
spin interacts with photons and its nuclear spin is used for storage.
This enables us to propose a simple but efficient MA-MDI-QKD scheme
resilient to memory errors and capable of beating, in terms of rate
and reach, existing QKD demonstrations. We also show how we can extend
this setup to a quantum repeater system, reaching, thus, larger distances. 
\end{abstract}
\maketitle

\section{Introduction}

Quantum repeaters (QRs) offer a fundamental solution to long-distance
quantum key distribution (QKD) \cite{Zoller_Qrepeater_98,DLCZ,Munro:NatPhot:2012,Repeater_review}.
The main building blocks needed for the first generations of QRs are
quantum memories (QMs). However, with the current technology, the
requirements on QMs are too demanding to allow high-rate key exchange
at long distances \cite{IEEE2, Muralidharan2016}. Memory-assisted measurement-device-independent
QKD (MA-MDI-QKD) is possibly the first step toward that end that can relax some of
the constraints on the QMs and, at the same time, allows to enhance
the rate-versus-distance behavior of a QKD system over a certain distance range \cite{Brus:MDIQKD-QM_2013, panayi}. It resembles a
single-node quantum repeater with QMs only in the middle site. The
performance of such a system depends considerably on the QMs in use.
Among the variety of QMs that can be used, nitrogen vacancy (NV) centers
in diamond embedded into cavities are promising candidates for a real
implementation of an MA-MDI-QKD setup \cite{NVcenterlopiparo,ScottQR,MAQKD_Delft}.
In \cite{NVcenterlopiparo}, the authors show that, by using the electron
spin states of two NV centers, such a setup can outperform the conventional
no-memory systems, as well as some other memory-assisted setups, in
a moderate-to-high coupling regime. Here, we propose a new configuration
that only uses one NV center. 
It relies on the nuclear spin of the NV center for storing quantum
states, while using its electron spin to interact with light. 
The simplicity of our scheme, combined with its superb noise resilience,
makes our setup suitable for implementation using a technology reachable
in the near future.

The performance of an MA-MDI-QKD setup depends strongly on how fast
the QMs in use can interact with light. MA-MDI-QKD can lead to practical advantages
when, compared to a no-memory system, the repetition rate is not too
slow. That would typically require a repetition rate on the order
of tens-to-hundreds of MHz. This condition is met by certain types
of QMs, whose interaction times can be on the order of nanoseconds. NV
centers, for instance, have an interaction time around 20~ns, which
makes them a possible candidate for MA-MDI-QKD.

In addition, there must be some heralding mechanism that announces
that the user's state has been stored into the QM. One way of doing
this is to teleport the user's state into the QM through a side Bell-state measurement (BSM). In particular, the state sent by each user
will interact with a photon previously entangled with certain internal
degrees of freedom of the NV center. Hence, upon a successful BSM,
the user's state will be teleported into the QM. The success of the
BSM on photons is heralded by a specific click pattern of the photodetectors
used in the BSM module.

When such a heralding method is used, the rate will be limited by
the time needed to entangle the photon with the QM. Therefore, the
entangling process is fundamental to determining the fastest repetition
rate that can be reached by such an MA-MDI-QKD setup. There are several
ways to entangle a photon with a QM depending on the type of memories
in use. The approach proposed in \cite{NVcenterlopiparo} relies on
using two NV centers in diamond, one for each user, as QM units. In
order to increase the probability of creating photon-QM entanglement,
the NV centers are also embedded into a one-sided microcavity. With
this cavity configuration, as well as with the assumption of a moderate
coupling regime, the authors present a setup that allows to teleport
the user's state into the electron spin of the NV center through the
so called double-encoding technique \cite{NVcenterlopiparo,RUS}.
For such a system, in \cite{NVcenterlopiparo}, the authors calculate
the secret key rate and compare it with that of other single-excitation
QMs, such as quantum dots, trapped ions and trapped atoms. They show
that NV centers embedded into microcavities outperform these other
QMs.

The results in \cite{NVcenterlopiparo}, while promising, may suffer
from the limited coherence time of the electron spin of the NV center
\cite{T2,NV_nuclearspin}, which makes the key rate of this system
to drop to zero at a distance around 500~km \cite{NVcenterlopiparo}.
Once other non-idealities, such as the additional background noise
from frequency converters \cite{MAQKD_SPS}, are accounted for, the
window over which the NV-based system outperforms the no-memory one
would even become narrower. It is therefore important to come up with
a system that has a wide window of opportunity, so that in practice
part of it can be exploited. For an NV center, such an enhanced performance
can be achieved by using the nuclear spin whose coherence time is
known to be much longer than that of the electron spin \cite{Bill_paper}.

Motivated by the possibility of using the nuclear spin of an NV center
as a storage unit, in this paper, we propose a new MA-MDI-QKD setup
that relies on {\em only} one NV center as physical memory. In
our setup, the electron spin is still being used to interact with
photonic systems. Here, we first load the NV center electron spin
with Alice's photon. Once a successful loading event takes place,
the electron spin state is transferred to the nuclear one. We then
proceed with loading the electron spin with Bob's photon. When the
electron spin is loaded for the second time, specific operations on
the electron and the nuclear spins will be performed to replicate
a {\em deterministic} full BSM operation creating, thus, a correlated
bit between the users.

For our proposed protocol, we calculate the secret key generation
rate, as the main figure of merit, and compare it with the fundamental
bound, as obtained in Ref.~\cite{Pirandola} and referred to by PLOB, hereafter, on the key rate of repeaterless
QKD systems. Our analysis accounts for major sources
of imperfection such as the decoherence and gate errors in the QM
as well as dark current in detectors and path loss. Moreover, we will
show how this scheme can be extended to a quantum repeater setup,
for which we estimate the longest secure distance possible.


The paper is structured as follows. In Sec. II we present our single-memory
MA-MDI-QKD scheme. In Sec. III, we describe our methodology for calculating
the secret key generation rate for the proposed protocol. We continue
by providing some numerical results in Sec. IV, before drawing our
conclusions in Sec. V.

\section{Single-memory MDI-QKD}


\begin{figure*}
\begin{centering}
\includegraphics[scale=0.07]{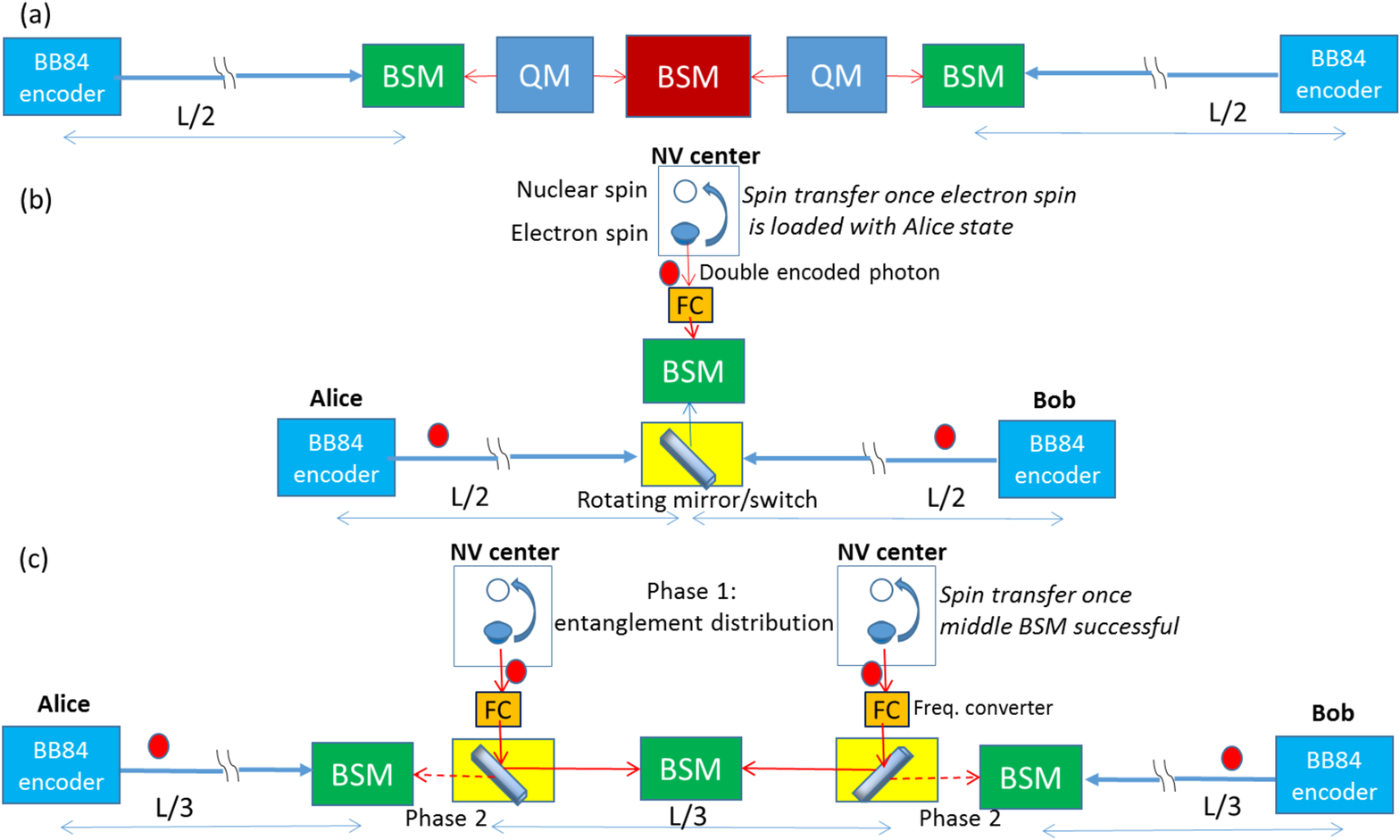} 
\par\end{centering}
\caption{\label{fig:single_memory_schemes}MA-MDI-QKD with (a) two non-heralding
physical memories, (b) a single memory with two qubits, and (c) its
extension to a three-leg quantum repeater system.}
\end{figure*}

\begin{figure}
\begin{centering}
\includegraphics[scale=0.6]{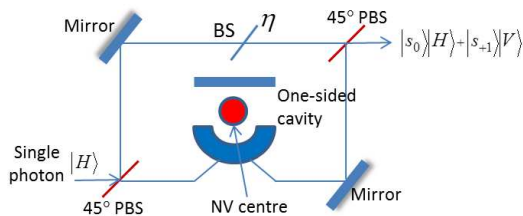} 
\par\end{centering}
\caption{\label{fig:DE_scheme}The double-encoding module proposed in \cite{NVcenterlopiparo}.
It allows to entangle a polarized photon with an NV center in a cavity.
This module will be used for transferring users' states into the electron
spin of the NV center.}
\end{figure}

Our single-memory MDI-QKD scheme belongs to the set of MA schemes
that use indirect heralding; see Fig. \ref{fig:single_memory_schemes}(a).
This set allows faster writing times as compared to certain directly
heralding memories \cite{panayi}. In Fig.~\ref{fig:single_memory_schemes}(a),
we first entangle a photon with some internal degrees of freedom of
the QM, and then interfere this photon and the one sent by the user
at a side-BSM. If the side-BSM is successful, the user's state has
ideally been teleported into the QM. In \cite{NVcenterlopiparo},
this required entangling operation is done by the double-encoding
module in Fig.~\ref{fig:DE_scheme}. This module relies on a cavity-based
NV center that, depending on its internal state, would impose a different
phase shift on an impinging single photon. As a result, an entangled
photon with the electron spin of the NV center can ideally be obtained.
We have summarized the details of this procedure in Appendix~\ref{AppA}.


Our proposed scheme in Fig.~\ref{fig:single_memory_schemes}(b) relies
on the same double-encoding scheme as in Fig.~\ref{fig:DE_scheme},
but it differs from the scheme in Fig.~\ref{fig:single_memory_schemes}(a)
in several ways. First, we have replaced the two physical QMs in Fig.~\ref{fig:single_memory_schemes}(a)
with only one NV center in Fig.~\ref{fig:single_memory_schemes}(b).
In Fig.~\ref{fig:single_memory_schemes}(b), both users send BB84
encoded photons at a similar repetition period, $T$, to the middle
site, at which, with the same period, a photon is double encoded with
the NV center electron spin. At this site, we first switch the photons
sent from, for instance, the left user (Alice) to the BSM module in
Fig.~\ref{fig:single_memory_schemes}(b) until a successful loading
occurs. That is, the state of Alice is teleported to the electron
spin. At this point, we transfer the electron spin state to the nuclear
one, and, in the meantime, rearrange the switch to now direct the
photons sent from the user on the right (Bob) to the BSM module. The
required optical switch here just needs to act like a rotating mirror,
and can be implemented using fast switching technologies \cite{bogoni2012photonic}.
Once we get the second successful loading event, the NV center contains
the state of Alice in its nuclear spin and that of Bob in its electron
spin. We just then need a BSM operation on these two systems to share
a key bit between Alice and Bob. The final BSM can be done deterministically, although with some possible errors
\cite{QCMC2016}, which we account for in our key rate analysis. The detailed description of the above steps is given in Appendix~\ref{AppB}.



An interesting feature of our proposed scheme is that by adding another
NV center to the scheme of Fig.~\ref{fig:single_memory_schemes}(b)
we can implement a three-leg quantum repeater setup, as shown in Fig.~\ref{fig:single_memory_schemes}(c),
with equidistant segments. This repeater system works in two phases.
First, we entangle the electron spins of the two NV centers in Fig.~\ref{fig:single_memory_schemes}(c),
which are separated by a distance $L_{0}=L/3$. This can be done by
double encoding a photon with each QM and then performing a BSM on
the two photons. Considering NV centers embedded in cavities, this
procedure is expected to succeed with a probability proportional to
the channel transmissivity. We then transfer the entangled state of
the electron spins to the nuclear spins and that will complete phase
1 of the protocol. In phase 2, we continue double encoding photons
with the electron spins, but now direct these photons to the side
BSM modules in Fig.~\ref{fig:single_memory_schemes}(c) using optical
switches. As soon as a successful side-BSM occurs, we can proceed
to perform a deterministic BSM on the nuclear and electron spin of
that QM as before. Note that, throughout both phases, users are sending
their encoded photons to the side BSM, but the middle sites do not
activate the side BSM modules until entanglement between the two QMs
is established. When both middle sites have swapped entanglement,
we are left with a shared key bit between the two end users. Note
that a similar but extended version of our three-leg quantum repeater
has been proposed in \cite{ScottQR}. They, however, use the double-heralding
technique for the initial entanglement \cite{NVexp2}, which is suitable
for NV centers without a cavity, but its rate scales with the product
of the channel transmissivity and the NV center coupling efficiency.
The latter is a limiting factor when cavity enhancement is not present
and would reduce the achievable rate in the system.



\section{Key rate analysis}

In this section the secret key generation rate of the proposed setup
of Fig. \ref{fig:single_memory_schemes}(b) is obtained under the
normal operation conditions when no eavesdropper is present. We will
also estimate the key rate of the quantum repeater scheme of Fig.~\ref{fig:single_memory_schemes}(c).
We assume that Alice and Bob use single photons in their encoders.
This is not a fundamental restriction but it provides a convenient
approach to compare memory-assisted schemes with the no-memory MDI-QKD systems. It is also possible to use decoy states, for which similar
margins of improvement over decoy-state no-QM systems are expected.
In \cite{panayi}, the total secret key generation rate, using the
efficient QKD protocol when ideal single photon sources are used by
the users and the $Z$ basis is more often used than the $X$ basis,
is lower bounded by the following expression 
\begin{equation}
R_{\mathrm{QM}}=\frac{R_{S}}{N_{L}\left(P_{A},P_{B}\right)+N_{r}}Y_{11}^{QM}(1-h(e_{11;X}^{\mathrm{QM}})-fh(e_{11;Z}^{\mathrm{QM}})),\label{eq:key_rate}
\end{equation}
where $P_{A}$ and $P_{B}$ are the probability of a successful side-BSM
on, respectively, Alice and Bob's side; $Y_{11}^{QM}$ is the probability
that the middle BSM is successful assuming that both memories are
loaded (in the Z basis); $e_{11;X}^{\mathrm{QM}}$ and $e_{11;Z}^{\mathrm{QM}}$
are, respectively, the quantum bit error rate (QBER) between Alice
and Bob in the $X$ and $Z$ basis when single photons are sent by
the users; $f$ is the inefficiency of error correction; $h(q)=-q\log_{2}q-(1-q)\log_{2}(1-q)$
is the binary entropy function; $R_{S}=1/T$ is the repetition rate;
$N_{L}$ is the average number of trials to load both memories, which
is given by $1/P_{A}+1/P_{B}$; and, $N_{r}=\lceil{{\tau_{r}}/{T}}\rceil$
is the number of additional rounds needed for reading and initializing
the QM, as well as any other required processing. The time corresponding
to these tasks is denoted by $\tau_{r}$.

In a real experiment, one needs to estimate the above parameters in order to use an appropriate level of error reconciliation and privacy amplification. Fortunately, in the limit of sufficiently long keys, all these parameters can be estimated accurately by conventional statistical techniques that correspond count rates to probabilities. If we use ideal single-photon sources, as we assume here, $Y_{11}^{\rm QM}$, $e_{11;X}^{\rm QM}$, and $e_{11;Z}^{\rm QM}$ can directly be estimated from the observed measurements. In the case of decoy-state encoding, we can use known techniques for bounding these parameters even in the finite-size key setting \cite{MDIQKD-Finite-key-Gaussian, curty2014finite, Finite-key-Chernoff-PRA}. Although we do not account for finite-size key effects in this comparative analysis, it is expected that MA-MDI-QKD, because of its better loss tolerance, would be less affected by this issue than its no-QM counterparts.
\subsection{Timing of the protocol}

\begin{figure}
\begin{centering}
\includegraphics[width=0.9\linewidth]{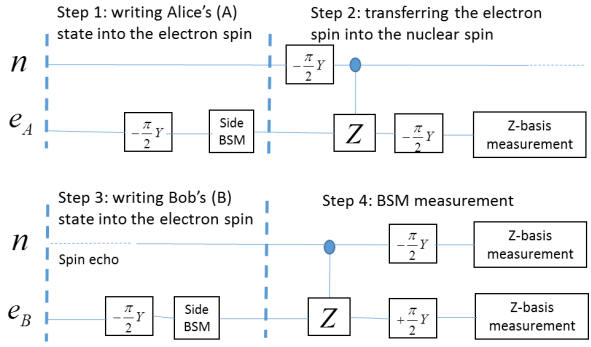} 
\par\end{centering}
\caption{\label{fig:Single_memory protocol}Schematic diagram for the sequential
operations in the protocol on each of the nuclear ($n$) and electron
($e$) spins. The subscript A (B) refers to the interaction of the
electron spin with Alice (Bob) photon. The detailed description of
each step and its required operations is given in Appendix~\ref{AppB}.}
\end{figure}

Figure \ref{fig:Single_memory protocol} schematically shows different
steps of our protocol and what operations each of the electronic and
nuclear spins would go through. In step 1, we try to load Alice's
photon. For each photon, we need to initialize the QM, which takes
$\tau_{{\rm init}}$, entangle a photon with it, taking $\tau_{{\rm int}}$,
and then do the side-BSM, taking $\tau_{M}$. Once we get to Bob's
photon (step 3), in addition to these operations, we also use spin
echo, with a time parameter $\tau_{{\rm dis}}$, to disentangle the
electron from the nuclear spin and to minimize any back action on
the stored state in the nuclear spin. The fastest we can repeat the
protocol then is given by $T=\tau_{\mathrm{init}}+\tau_{\mathrm{int}}+\tau_{{\rm dis}}+\tau_{{\rm M}}=\tau_{w}$.
In principle, this can be done faster for Alice's photon at $T=\tau_{w}-\tau_{{\rm dis}}$,
but we ignore this possible advantage in our analysis.

In addition to capturing Alice and Bob's photons, in Fig.~\ref{fig:Single_memory protocol},
we also need to transfer the state of electron spin to the nuclear
spin (step 2), taking a time denoted by $\tau_{{\rm swap}}$, and
finally do a deterministic BSM on the two spins (step 4) taking $\tau_{{\rm BSM}}$.
In which case, $\tau_{r}=\tau_{{\rm swap}}+\tau_{{\rm BSM}}+\tau_{{\rm init}}$.


\subsection{Error Analysis}

In order to calculate the elements of Eq.~\eqref{eq:key_rate}, in our simulation, we use
the same approach as the one used in \cite{NVcenterlopiparo} with
regard to the modeling of the decoherence, loss, dark count and detection
efficiency. In addition to those errors, here, we particularly account
for the errors in each of the logic gates of Fig.~\ref{fig:Single_memory protocol}
by using a depolarizing channel inspired model. In particular, for a single-qubit
logic gate $R_{k}$ on spin $k=e,n$, we assume that the joint state
of electron-nuclear spins $\rho_{en}$ undergoes the following transition:
\begin{eqnarray}
\rho_{en}\rightarrow && (1-p_k)R_k\rho_{en}R^{T} \nonumber\\
&&+\frac{p_k}{3}\left(X_k\rho_{en}X_k+Y_k\rho_{en}Y_k+Z_k\rho_{en}Z_k\right),
\label{eq:dep_channel}
\end{eqnarray}
where $X_k$, $Y_k$, and $Z_k$ are the Pauli matrices of spin $k$ and $p_k$ is the corresponding depolarization parameter.  

The only two-qubit gate in Fig.~\ref{fig:Single_memory protocol} is the controlled-Z (CZ) gate, whose operation is modeled as follows:
\begin{eqnarray}
\rho_{en}\rightarrow && (1-p_{\rm CZ})O_{\rm CZ} \rho_{en} O_{\rm CZ}  \nonumber \\
&&+ p_{\rm CZ}/3(X_e X_n \rho_{en} X_e X_n)  \nonumber\\
&&+ p_{\rm CZ}/3(Y_e Y_n \rho_{en} Y_e Y_n)  \nonumber\\
&&+ p_{\rm CZ}/3(Z_e Z_n \rho_{en} Z_e Z_n) ,
\end{eqnarray}
where $O_{\rm CZ}$ represents the ideal CZ gate operation. In principle, one can also include additional error terms for different electron-nuclear Pauli operators. But, numerically, we find the above model sufficiently accurate for the purposes pursued in this paper. 

Using the above models, we can obtain the state of the system at any
step in Fig.~\ref{fig:Single_memory protocol}. The derivations are
cumbersome and have mostly been done by the software Maple. In short,
for each scheme, we first obtain the state of the QMs once the user's
state is loaded to them. At this stage, we also find $P_{A}$ and
$P_{B}$ by including the number of rounds lost due to the deadtime
as explained in \cite{NVcenterlopiparo}. Finally, we calculate the
remaining terms in Eq.~(\ref{eq:key_rate}), i.e., $Y_{11}^{QM},$
$e_{11;X}^{\mathrm{QM}}$, and $e_{11;Z}^{\mathrm{QM}}$.

\subsection{Three-leg Repeater}

\label{Sec:RateRep} In principle, we can use the above detailed analysis
to calculate the rate for the repeater setup in Fig.~\ref{fig:single_memory_schemes}(c).
But, as will be shown later, our scheme is quite resilient to existing
gate errors. We therefore only provide a rough estimate of the key
rate for the repeater scheme of Fig.~\ref{fig:single_memory_schemes}(c).
To that end, we calculate the average time $T_{{\rm rep}}$ that it
takes to generate a raw key bit between Alice and Bob. This parameter
includes the average time $T_{{\rm ent}}$ that it takes to entangle
the two electron spins, the time to transfer the electron spins to
nuclear spins, perform BSM operations on the two, and initialize the
QMs again, whose sum we denoted earlier by $\tau_{r}$, as well as the
time $T_{{\rm load}}$ for loading electron spins with Alice and Bob
photons. In this case, we have 
\begin{equation}
T_{\mathrm{rep}}=T_{{\rm ent}}+\tau_{r}+T_{{\rm load}},\label{eq:rate_rep}
\end{equation}
where 
\begin{equation}
T_{{\rm ent}}\approx T_{0}/P_{{\rm ent}}
\end{equation}
with $T_{0}=(L/3)/c$ being the transmission delay in the middle link
(with $c$ being the speed of light) and 
\begin{equation}
\label{eq:Pent}
P_{{\rm ent}}=(1/2)(\eta\eta_{s}\eta_{c}\eta_{d})^{2}e^{-(L/3)/L_{{\rm att}}}
\end{equation}
being the success probability of a polarization-based entanglement
distribution scheme where photons are entangled with QMs and a probabilistic
BSM will be performed on these photons in the middle of the link.
Here, $\eta$ is the efficiency of the double encoding module of Fig.~\ref{fig:DE_scheme},
$\eta_{s}$ is the efficiency of the single-photon source used in
that module, $\eta_{d}$ is the detector efficiency, $\eta_{c}$ is
the frequency converter efficiency, and $L_{{\rm att}}$ is the attenuation
length of the channel. In Eq.~\eqref{eq:Pent}, $\eta \eta_s \eta_c$ represents the probability of entangling a photon with the electron spin, $\eta_d^2/2$ represents the BSM success probability, and $\exp(-L/3/L_{\rm att})$ represents the channel loss. Finally, in Eq.~\eqref{eq:rate_rep}, 
\begin{equation}
T_{{\rm load}}\approx(3/2)T/P_{A},
\end{equation}
which accounts for the average time needed to load both photons assuming that $P_A=P_B$ \cite{panayi}. Using
the above timing calculations, we then estimate the secret key rate
of the setup of Fig.~\ref{fig:single_memory_schemes}(c) by $R_{{\rm rep}}=1/T_{{\rm rep}}$. Note that this is an optimistic estimate of the rate in which memory errors are not accounted for.


\section{Numerical results}

\begin{table}
\begin{centering}
{\footnotesize{}{}}%
\begin{tabular}{|c|c|}
\hline 
\textcolor{black}{\scriptsize{}{}Entangling efficiency, $\eta$}{\scriptsize{} } & \textcolor{black}{\scriptsize{}{}$0.9$}{\scriptsize{}{} }\tabularnewline
\hline 
{\scriptsize{}{}Cooperativity, C}  & {\scriptsize{}{}50}\tabularnewline
\hline 
{\scriptsize{}{}Single photon source efficiency, $\eta_{s}$}  & {\scriptsize{}{}0.72}\tabularnewline
\hline 
\textcolor{black}{\scriptsize{}{}Frequency conversion efficiency,
$\eta_{c}$}{\scriptsize{}{} }  & \textcolor{black}{\scriptsize{}{}0.68}{\scriptsize{}{} }\tabularnewline
\hline 
\textcolor{black}{\scriptsize{}{}Detector efficiency, $\eta_{d}$}{\scriptsize{}{}
}  & \textcolor{black}{\scriptsize{}{}0.93}{\scriptsize{}{} }\tabularnewline
\hline 
\textcolor{black}{\scriptsize{}{}Dark count rate \cite{dark_count}}{\scriptsize{}{}
}  & \textcolor{black}{\scriptsize{}{}1 cps}{\scriptsize{}{} }\tabularnewline
\hline 
\textcolor{black}{\scriptsize{}{}Attenuation length, $L_{{\rm att}}$}{\scriptsize{}{}
}  & \textcolor{black}{\scriptsize{}{}25 km}{\scriptsize{}{} }\tabularnewline
\textcolor{black}{\scriptsize{}{}Speed of light in fiber, $c$}{\scriptsize{}{}
}  & \textcolor{black}{\scriptsize{}{}$2\times10^{8}$ m/s}{\scriptsize{}{} }\tabularnewline
\hline 
{\scriptsize{}{}Initialization time, $\tau_{{\rm init}}$ }  & {\scriptsize{}{}11.5 ns }\tabularnewline
\hline 
{\scriptsize{}{}Interaction time, $\tau_{{\rm int}}$ }  & {\scriptsize{}{}10 ns }\tabularnewline
\hline 
{\scriptsize{}{}Verification time, $\tau_{{\rm M}}$ }  & {\scriptsize{}{}1 ns }\tabularnewline
\hline 
{\scriptsize{}{}Swap time, $\tau_{{\rm swap}}$}  & {\scriptsize{}{}1.1 $\mu\mathrm{s}$ }\tabularnewline
\hline 
{\scriptsize{}{}Deterministic BSM time, $\tau_{{\rm BSM}}$}  & {\scriptsize{}{}1.5 $\mu\mathrm{s}$ }\tabularnewline
\hline 
{\scriptsize{}{}Error probability of the $\pm\frac{\pi}{2}Y$ }  & {\scriptsize{}{}$10^{-3}$}\tabularnewline
{\scriptsize{}{}gate for the electron spin, $p_{e}$}  & \tabularnewline
\hline 
{\scriptsize{}{}Error probability of the CZ gate, $p_{{\rm CZ}}$}  & {\scriptsize{}{}$2\times10^{-4}$}\tabularnewline
\hline 
{\scriptsize{}{}Error probability of the $-\frac{\pi}{2}Y$ }  & {\scriptsize{}{}$10^{-3}$}\tabularnewline
{\scriptsize{}{}gate for the nuclear spin, $p_{n}$}  & \tabularnewline
\hline 
\end{tabular}
\par\end{centering}

 {\footnotesize{}{}\caption{\label{tab:Nominal-values-used}Nominal values used in our numerical
results. }
}{\footnotesize \par}

\end{table}

In this section, we compare the rate of our proposed single-memory
MDI-QKD scheme with the PLOB bound of a repeaterless QKD system reported
in Ref.~\cite{Pirandola}. For a pure-loss channel with a total transmissivity $\eta_T$, the maximum achievable secret key rate per transmitted pulse in a repeaterless system is given by $\log_2(1-\eta_T)$. In our simulations, we assume that $\eta_T = \exp(-L/L_{\rm att}) \eta_d$. We multiply this bound by relevant clock rates in no-QM systems to obtain a bound on the total key rate. We also estimate
the rate of the three-leg quantum repeater scheme that relies on our
proposed scheme of Fig. \ref{fig:Single_memory protocol}(c). The
nominal values used in our numerical analysis are summarized in Table
\ref{tab:Nominal-values-used}. These values are taken from the state-of-the-art
technologies for various quantum devices. The time and error parameters
of the NV centers is taken from the rigorous analysis in \cite{gates_errors}.
In our analysis, we also consider the use of frequency converters
to allow the interaction between a QM-driven photon and the telecom
photon sent by the user. We model this as an additional source of
loss, which modifies the efficiency of the side-BSM detectors \cite{upconversion,up_conv1}.

\subsection{Single-memory MDI-QKD}

\begin{figure}
\begin{centering}
\includegraphics[width=.8 \linewidth]{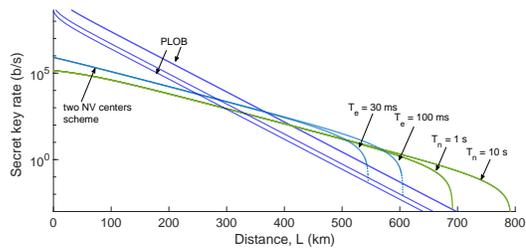}
\par\end{centering}
\caption{\label{fig:key_rate_vs_no_mem}Secret key generation rates versus
distance for our proposed single-memory MDI-QKD scheme and its comparison
with a no-QM system (PLOB) driven at 100 MHz (lower curve) and 1~GHz
(upper curve) as well as the two-QM setup proposed in \cite{NVcenterlopiparo}.
In the latter case, the coherence time of the electron spin, $T_{e}$,
ranges from 30 ms to 100 ms. $T_{n}$ refers to the coherence time
of the nuclear spin, which is the relevant time constant for the setup proposed here.}
\end{figure}

Figure \ref{fig:key_rate_vs_no_mem} compares the secret key generation
rate of our proposed single-memory MDI-QKD scheme with two different
alternatives: the PLOB bound and the two-QM scheme in \cite{NVcenterlopiparo}.
The rate of our proposed scheme outperforms the PLOB bound at a distance
around 400~km if we assume that the repeaterless system is driven by an ideal single-photon source with a pulse rate of 1 GHz. In Fig.~\ref{fig:key_rate_vs_no_mem}, we have also included a PLOB curve when the pulse rate of the ideal source is 100~MHz. This somehow replicates the practical case where, instead of single-photon sources, one may use the decoy-state technique. For decoy-state encoding, when the average number of photons per signal state is 0.5, only 30\% of the time we generate single-photon states. If one accounts for the percentage of the time that the signal state, rather than decoy states, may be used, one can argue that only about 10\%-20\% of the pulses sent will carry single-photon states. That is even if we send 1 Gpulse/s, only 100-200 Mpulse/s would carry single-photon information. The lower PLOB curve in Fig.~\ref{fig:key_rate_vs_no_mem} shows this scenario by calculating the rate for an ideal single-photon source with a pulse rate of 100~MHz. In this case, the cross-over distance is around 300~km.

 Our scheme offers a slightly lower rate than that of the two-NV-center scheme in Ref.~\cite{NVcenterlopiparo} at short-to-medium distances, but can extend the maximum security distance by a few hundred kilometers. The reason for the drop in rate is partly due slower repetition rates. In our scheme, we consider doing spin echo in every round, which takes around $\tau_{{\rm dis}}=20$~ns, and that would almost reduce the repetition rate by a factor of two. We instead do the middle BSM deterministically, which should buy us a factor of two if we had no errors in our operations. But, then, instead of loading both memories in parallel as in \cite{NVcenterlopiparo}, we do it sequentially, and that is why the starting rate for the scheme of Ref.~\cite{NVcenterlopiparo} is a bit higher. As we get to longer and longer distances, the decoherence issue with electron spin states, in the scheme of Ref.~\cite{NVcenterlopiparo}, would kick in, whereas our single-QM scheme, which relies on nuclear spins, is still resilient to decoherence errors. As a result, the maximum security distance considerably improves, and that gives us some more immunity against other sources
of noise that may exist in a realistic setup. Moreover, in terms of resources, the scheme of Fig.~\ref{fig:single_memory_schemes}(b)
uses almost half of major devices needed in the scheme of Ref. \cite{NVcenterlopiparo}. If we consider the normalized rate per resources used, as a figure of merit, then the two schemes offer similar rates, but the one that
uses nuclear spins cover longer distances.

\begin{figure}
\begin{centering}
\includegraphics[width=0.9\linewidth]{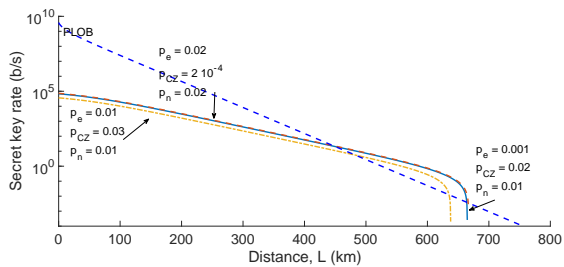} 
\par\end{centering}
\caption{\label{fig:rate_diff_p}Secret key generation rates versus distance
for our proposed single-memory MDI-QKD scheme and its comparison with
no-QM MDI-QKD driven at 1~GHz for different values of error probabilities
associated with the logic gates of Fig.~\ref{fig:Single_memory protocol}. }
\end{figure}

Another source of error that may limit the performance of our system
is the errors associated with the logic gate operations of Fig.~\ref{fig:Single_memory protocol}.
In \cite{gates_errors}, the authors analytically estimate the errors
caused by these operations for different values of the pump strength.
However, in a real experimental setup they could differ from the results
obtained in \cite{gates_errors}. Therefore, we also estimate the
highest error tolerable by our system within which it is still possible
to have an improvement over the PLOB bound. Figure~\ref{fig:rate_diff_p}
shows several extreme cases where $p_{e}$, $p_{n}$, and $p_{{\rm CZ}}$
take rather large values on the order of 0.01. It can be seen that,
at $T_{n}=1$ s, even such high error terms will kick in after 600
km of channel length. Note that the tolerable error rate is over one
order of magnitude higher than the expected nominal values. That would
give us assurance that such a setup, once extended to a quantum repeater
setting, can tolerate multiple rounds of entanglement swapping without
any need for purification.

In addition to gate errors and memory decoherence, there are other practical aspects that one needs to deal with in a realistic scenario. For instance, the single-photon source used in the middle of the link must have very low two-photon emission rates otherwise, these additional photons may cause errors in the system \cite{IEEE1}. In our case, the middle QMs are driven by NV-center-based single-photon sources that have a very similar structure to the QMs themselves. The chance of multiple-photon emission is then naturally kept low. Note that the multiple photon terms generated by the users, in a decoy-state scenario, would not cause much problems as they go through a lossy channel. Similarly, one should be aware of the background noise generated by external sources, such as classical channels propagating over the same fiber medium as the quantum ones \cite{crosstalk2016}, or by the frequency converters. By proper design and filtering \cite{OFDM-QKD_SPIE-Photon2016}, this should be achievable in practice. Our scheme also additionally needs a $2\times2$ optical switch to swap between two users. For such a switch one should consider the insertion loss as well as the switching time. The latter issue is less of a problem in our case, as, in the limit of long distances, the time between two consecutive photons surviving the path loss is rather long, and that would alleviate the requirements on the switching time. In such a case, for such a small-size switch, the insertion loss could also be sufficiently low. In our numerical results, we have neglected the switching insertion loss. Finally, in practice, it is rarely the case that the distance between the user nodes and the middle node are identical. While this may cause us to deviate from the optimal performance, the effect can easily be modeled within our framework. The result is not expected to be much different from that of an asymmetric no-QM MDI-QKD setup.

\subsection{Three-leg Repeater}

Now that we establish that the errors arising from our logic gates
is very low, we can reliably use the estimate in Sec.~\ref{Sec:RateRep}
to calculate the key rate of the quantum repeater setup of Fig. \ref{fig:single_memory_schemes}(c).
For simplicity, we have also ignored the effect of dark count. For
a dark count rate of 1 cps, and assuming pulse widths on the order
of nanoseconds, the dark count will become important when $L/3$ is
comparable to 450~km. We should therefore be able to cover distances
up to around 1400~km using this simple repeater setup. If other sources
of background, such as the Raman noise from classical channels or
the frequency converters, kick in, then the corresponding maximum length
would be reduced. 
\begin{figure}
\begin{centering}
\includegraphics[width=0.8\linewidth]{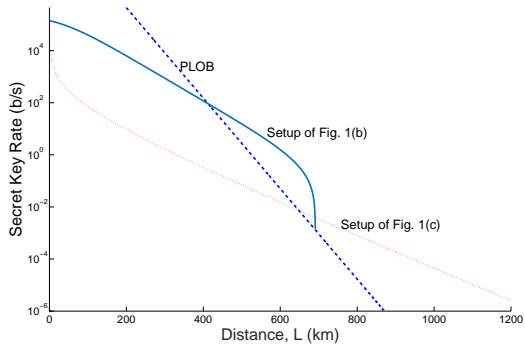} 
\par\end{centering}
\caption{\label{fig:repeater}Secret key generation rates versus distance for the quantum repeater
setup of Fig.~\ref{fig:single_memory_schemes}(c) and our proposed
single-memory MDI-QKD scheme at $T_{n}=1$ s.}
\end{figure}

Figure \ref{fig:repeater} shows the comparison between
the single-memory MDI-QKD scheme at $T_{n}=1$ s and its quantum repeater
extension of Fig.~\ref{fig:single_memory_schemes}(c). It can be
seen that the repeater setup outperforms the PLOB bound around 600~km,
where the rate is already very low at around $R_{{\rm rep}}\sim10^{-2}/\mathrm{b/s)}$.
In order to increase the rate, we might consider a multiple memory
configuration as proposed in \cite{Razavi.Lutkenhaus.09,Razavi.Amirloo.10, meas_op, IEEE2} or use more nesting levels
as in Ref. \cite{ScottQR}. In \cite{ScottQR}, the authors calculate
the secret key rate versus the distance for different values of the
gate efficiency with and without entanglement distillation. In order
to reach long distances, they however need gate efficiencies around
0.99. Such high efficiencies may be possible if one uses the cavity
setting that we have assumed for the NV centers. In any case, the
general repeater setup still seems to be very hard to implement with
current technologies.

\section{Conclusions}

In this paper, we presented a new MA-MDI-QKD system, which relied
on only one physical memory: an NV center embedded in a small-volume
optical cavity. We calculated the secret key rate for such a system
and we compared it with the fundamental bound for a repeaterless QKD
system as well as with the two-NV-center system proposed in \cite{NVcenterlopiparo}.
In our new system, for storage, we relied on the nuclear spin of the
NV center, which could have a coherence time as long as 10 seconds.
For such a value of coherence time, we showed that the rate of our
system can reach longer distances compared to the scheme proposed
in \cite{NVcenterlopiparo}, while the normalized rate per NV center module used was about the same. We also determined the highest tolerable error probability for each logic gate operation required to extract a secret key
when we use this scheme. We compared these threshold error rates with
what is expected from these devices, and showed that we had sufficient
room to accept additional error in the system.

While MA-MDI-QKD is a good approach to enhance the rate-versus-distance
behavior of a QKD system, in order to achieve longer distances, we
need a quantum repeater setup. Therefore, we showed how we could extend
our single-memory system to a simple quantum repeater setup by adding
one more memory and splitting the channel into three equidistant elementary
links. For such a setup, we estimated the longest distance achievable
by considering the time intervals required for each single operation.
We showed that we could reach a total distance of about 1400 km. After
this distance the dark count rate would become dominant not allowing
to extract a secret key. Despite the longer distance the quantum repeater
setup allowed us to reach, its rate was very low. To address such
an issue, we should consider a multiple memory configuration. In this
way we can speed up the probability of storing the users' states into
the memories. Nevertheless, the single-memory scheme presented in
this paper offered a simple implementation, which could be extended
to a quantum repeater system in the future.

\section*{Acknowledgments}
This work was funded by the JSPS International Research Fellowship.
This work was partly funded by the UK's EPSRC Grant EP/M013472/1 and EPSRC
Grant EP/M506951/1, and the EU's H2020 programme under the Marie Sk\l{}odowska-Curie project QCALL (GA 675662).
WJM acknowledges  support through a grant from the John Templeton Foundation. The opinions expressed in this publication are those of the author(s) and do not necessarily reflect the views of the John Templeton Foundation.

\appendix

\section{Double encoding scheme}

\label{AppA} In this Appendix, we review the entangling scheme used
in \cite{NVcenterlopiparo}, shown in Fig.~\ref{fig:DE_scheme},
known as the double-encoding module. The double-encoding scheme is
used to entangle a photon with the electron spin states of the NV
center. To this end, the NV center is embedded into a one-sided cavity,
whose effective reflectivity is affected by the internal state of
the NV center. The idea of conditional reflectivity has already been
proposed in \cite{duan_kible} for a trapped atom system. In particular,
in \cite{NVcenterlopiparo}, the authors show that when the NV center
is in the electron spin state 0, $|s_{0}\left\rangle \right.,$ then
the incoming photon to the module of Fig. \ref{fig:DE_scheme} will
be reflected off the cavity. When the NV center is in the electron
spin state +1, $|s_{+1}\left\rangle \right.,$ then the photon will
also be reflected but it will acquire a $\pi$ phase. This implies
that in both cases the photon will be reflected but with a different
phase shift.

The module of Fig. \ref{fig:DE_scheme} ideally works as follows.
First the NV center is initialized into the state $|\Psi_{in}\left\rangle =\left(|s_{0}\left\rangle \right.+|s_{+1}\left\rangle \right.\right)/\sqrt{2}\right..$
Then, we generate an $H$-polarized single photon and send it through
a $+45^{\circ}$ polarizing beam splitter (PBS). We can generate such
a single photon by driving a specific transition in another cavity-NV-center
pair \cite{NVcenterlopiparo}. In Fig. \ref{fig:DE_scheme}, the $+45^{\circ}$-polarized
component of this single photon interacts with the NV center, resulting
in the joint state $|D\left\rangle _{s}\right.\left(|s_{0}\left\rangle \right.-|s_{+1}\left\rangle \right.\right)/\sqrt{2}$,
where $|D\rangle=\frac{1}{\sqrt{2}}\left(|H\rangle+|V\rangle\right)$.
The photonic modes $r$ and $s$ are then recombined at a second $+45^{\circ}$
PBS, which will result in the following output state 
\begin{equation}
|\Psi_{2}\left\rangle \right.=\frac{1}{\sqrt{2}}\left(|H\left\rangle \right.|s_{0}\left\rangle \right.+|V\left\rangle \right.|s_{+1}\left\rangle \right.\right).\label{eq:double-enc}
\end{equation}

In deriving Eq.~(\ref{eq:double-enc}), we have made the assumption
that the reflection coefficients, in the two cases of $|s_{0}\rangle$
and $|s_{+1}\rangle$ states, has the same magnitude of 1. However,
for finite values of the cooperatively C the two coefficients may
have different values, leading to a deviation from the ideal entangled
state in Eq.~(\ref{eq:double-enc}). As a consequence, this will
cause an imbalance between the two legs of the interferometer in Fig.
\ref{fig:DE_scheme}. We can fix this by adding a beam splitter with
transmissivity $\eta$ in the $r$ branch. The value of $\eta$ will
be chosen accordingly to account for different sources of loss in
the $s$ branch. In this case, the generated state by our double-encoder
will become \cite{NVcenterlopiparo} 
\begin{equation}
\rho_{\mathrm{NV-P}}=\eta|\Psi_{2}\left\rangle \right.\left\langle \right.\Psi_{2}|+(1-\eta)|\mathbf{0}\left\rangle \negthinspace\negthinspace\right.{}_{\mathrm{PP}}\hspace{-1mm}\left\langle \right.\mathbf{0}|\otimes I_{\mathrm{NV}}^{'},\label{eq:den_mat_sec_scheme}
\end{equation}
where $I_{\mathrm{NV}}^{'}=(|s_{0}\rangle\langle s_{0}|+|s_{+1}\rangle\langle s_{+1}|)/2$.

The output state of Eq. (\ref{eq:den_mat_sec_scheme}) will teleport
the state sent by the user into the QM in most of the cases when the
side-BSM is successful. The vacuum term will introduce a small error
that is proportional to the dark count. However, this error will not
be relevant as we will show in our calculation of the secret key rate.

Another practical assumption is considering a strong coupling regime
for the NV center embedded into the cavity. To this aim we consider
a cooperativity $C=50.$ In \cite{NVcenterlopiparo}, the authors
show that for lower values of C, the reflection coefficients of the
cavity strongly depend on the state of the NV center. On one hand,
as C decreases, when the NV center state is $|s_{0}\left\rangle \right.$
the reflection coefficient decreases as well and, on the other hand,
when the NV center state is $|s_{+1}\left\rangle \right.$ the reflection
coefficient approaches 1. In this case, as C decreases, the key rate
decreases as well. However, it is still possible to extract a secret
key for a value of C as low as 1.22 \cite{NVcenterlopiparo}. For
the sake of simplicity, in our calculation, we assume that the NV
center is in the strong coupling regime so we can use Eq. \ref{eq:den_mat_sec_scheme}
as output density matrix describing the double-encoded state of the
polarized photon with the NV center.

At the beginning of each round, before performing the double encoding
operation discussed above, we have to initialize the NV center in
state $|\Psi_{in}\rangle$. The initialization process can be performed
using the double-encoding module of Fig. \ref{fig:DE_scheme}. In
every round, we send an {$H$-polarized} single photon to the NV-center-cavity
module, and measure the polarization of the output photon in $|H\rangle$
and $|V\rangle$ basis. Depending on which photodetector will click,
we could infer the corresponding state of the NV center, i.e $|s_{0}\rangle$
or $|s_{+1}\rangle$. Then we apply the relevant rotation to initialize
the NV center in $|\Psi_{in}\rangle$. The above procedure includes
the double encoding operation and a rotation. The time for the double-encoding
operation is given by the time needed for the photon to interact with
the NV center, $\tau_{\mathrm{int}}$, i.e. the interaction time,
which takes roughly 10 ns. The rotations on the electron spin can
be driven by using a microwave driving field perpendicular to the
NV center axis without the same drive also affecting the nuclear spin
\cite{gates_errors}. In \cite{gates_errors}, the authors analytically
calculated the timing and the error associated to the rotations required
to initialize the NV center. They depend on the pump strength of the
driving field \cite{gates_errors}. In our work, we assume that the
pump strength is 375 MHz, which will rotate the electron spin in roughly
1.5 ns. Therefore, adding up both these operations we get 11.5~ns,
which corresponds to the initialization time, $\tau_{\mathrm{init}}$,
in our protocol.

In the above procedure, if we get no click, then we consider that
the initialization process has failed. This can happen for several
consecutive rounds, which indicates that the memory is in a deadtime
period \cite{NVcenterlopiparo}. During this period, the NV center
is in certain metastable states, which will decay to any of $|s_{0}\rangle$
and $|s_{\pm1}\rangle$ states \cite{NVcenterlopiparo}. Since $|s_{-1}\rangle$
does not correspond to any desired state, during the deadtime, we
swap states $|s_{0}\rangle$ and $|s_{-1}\rangle$ in every initialization
round to avoid the possibility that the NV center stays in the state
$|s_{-1}\rangle$ for ever.

\section{Single-memory MDI-QKD protocol}

\label{AppB} Figure \ref{fig:Single_memory protocol} shows the steps
containing all the required logic gates that must be applied to the
electron and nuclear spin of the NV center in order to extract a secret
key in the scheme of Fig.~\ref{fig:single_memory_schemes}(b). The
upper line in each section of Fig. \ref{fig:Single_memory protocol}
refers to the gate operations applied to the nuclear (n) spin and
the lower line refers to the gate operations applied to the electron
($e_{A},$ $e_{B}$) spin of the NV center. The subscripts A and B
of the electron spin refer to the event of storage of the electron
spin with Alice's and Bob's state, respectively.

The single-memory MDI-QKD protocol works as follows, see Fig. \ref{fig:Single_memory protocol}. 

{\bf Step 1, Alice Teleportation:} The first step consists of a $-\frac{\pi}{4}$ rotation around the
$Y$ axis of the electron spin of the NV center, which will initialize
the electron spin into the state $|\Psi_{in}\rangle$, and a side-BSM,
which, if successful, will project the user's state into the electron
spin. These two operations are represented by the $-\frac{\pi}{2}Y$
gate and Side-BSM in Fig.~\ref{fig:Single_memory protocol}. Rotations
on the electron spin can be implemented by using a microwave driving
field perpendicular to the NV center as explained in \cite{gates_errors}.

{\bf Step 2, Spin Transfer:} Once the electron spin has been written with the user's state, we
perform a $-\frac{\pi}{2}Y$ rotation on the nuclear spin, which will
create the state $|n_{+}\rangle=\frac{1}{\sqrt{2}}\left(|\uparrow\left\rangle \right.+|\downarrow\left\rangle \right.\right),$
where $|\uparrow\left\rangle \right.$ and $|\downarrow\left\rangle \right.$
are, respectively, the up and down nuclear spin states in the Z-basis.
After that, we let the nuclear spin interact with the electron spin
through the always-on hyperfine interaction with Hamiltonian $H_{\mathrm{eff}}=\hbar A_{\mathrm{net}}|s_{1}\rangle \langle s_{1}|\otimes|\uparrow\rangle \langle\uparrow|$,
where $A_{\mathrm{net}}$ is the coupling strength as explained in the supplementary
material of Ref.~\cite{Bill_paper}. The hyperfine interaction provides
a route to entangling the spins without resorting to driving fields
or varying the magnetic fields dynamically \cite{gates_errors}. In
fact, if we assume that the electron spin is in state $|\Psi_{e}\rangle=\alpha|s_{0}\rangle +\beta|s_{1}\rangle$, where $\alpha$ and $\beta$ are arbitrary coefficients corresponding to the state sent by the user, the joint state of the nuclear-electron system after the nuclear-spin rotation will be given by
\begin{equation}
\label{Eq:ne}
|\psi\rangle_{ne}=|n_{+}\rangle|\Psi_{e}\rangle.
\end{equation}
If we now let this state evolve according to $H_{\mathrm{eff}}$, we obtain
\begin{equation}
|\psi\rangle_{ne}={\alpha}|s_{0}\rangle|n_{+}\rangle+{\beta}|s_{1}\rangle(|\downarrow\rangle +e^{iA_{\mathrm{net}}t}|\uparrow\rangle )/\sqrt{2}.\label{eq:ent_n_el}
\end{equation}
At $ t = \pi / A_{\mathrm{net}}$, which roughly corresponds to $t = 165$~ns, the above state will become
\begin{equation}
|\psi\rangle_{ne}={\alpha}|s_{0}\rangle|n_{+}\rangle+{\beta}|s_{1}\rangle|n_{-}\rangle,
\end{equation}
which represents the CZ operation on the initial state in Eq.~\eqref{Eq:ne} with the nuclear spin as the control qubit. Now, in order to transfer the electron spin into the nuclear spin, we first rotate the electron
spin by another $-\frac{\pi}{2}Y$ operation obtaining 
\begin{equation}
|\psi\rangle_{ne}=[{\alpha}(|s_{0}\rangle+|s_{1}\rangle)|n_{+}\rangle+{\beta}(|s_{0}\rangle-|s_{1}\rangle)|n_{-}\rangle]/\sqrt{2}.
\label{eq:ent_n_el2}
\end{equation}
If we now measure the electron spin by using the double-encoding
procedure described in Appendix~\ref{AppA}, which corresponds to a Z-basis measurement
gate in Fig.~\ref{fig:Single_memory protocol}, the state of Eq.~\eqref{eq:ent_n_el2}
will become 
\begin{equation}
|\psi\rangle_{ne}=\alpha|n_{+}\rangle \pm \beta|n_{-}\rangle,
\end{equation}
where the sign depends on the outcome of the measurement.

The user's state is now stored into the nuclear spin. Note that the nuclear state undergoes a decoherence process, which has been taken into account in our calculation. However, in this case, we can rely on much longer
coherence times as compared to the scheme of \cite{NVcenterlopiparo},
due to the longer nuclear spin coherence time. 

{\bf Step 3, Bob Teleportation:} Now the other user, Bob, repeatedly tries to store his state into the electron
spin with the same procedure as in step 1. The only difference is
that we now have to do spin echo in every round to preserve the state
of the nuclear spin and prevent a back action on the nuclear spin
due to external interactions with the electron spin. 

{\bf Step 4, Final BSM:} Finally, in step 4, we perform a CZ gate and an X-basis measurement
on both electron and nuclear spins as shown in Fig.~\ref{fig:Single_memory protocol}.
As explained in step 2, the X measurements are done by rotating the
spins and then performing a Z measurement. This is equivalent to performing
a full BSM in order to create a correlated bit between Alice and Bob.

Depending on which basis and state Alice and Bob pick, we can have
different possible output states right before the final Z-basis measurements,
in step 4, on nuclear and electron spins. These states are summarized
in Table \ref{tab:outcomes}. Based on this table, in the Z-basis,
Alice and Bob, only need to account for the result of measurement
on the nuclear spin. A spin up means that they have both got similar
bits, and a spin down implies the other case. If they have both chosen
the X basis, then the electron spin would be in $|s_{0}\rangle$ if
Alice and Bob share the same bit, and in $|s_{+1}\rangle$ if they
have complementary bits.

\begin{table}[t]
\begin{centering}
\begin{tabular}{|c|c|}
\hline 
Z-basis  & X-basis\tabularnewline
\hline 
\end{tabular}
\par\end{centering}
\begin{centering}
\begin{tabular}{|c|c|c|c||c|c|c|c|}
\hline 
Alice  & Bob  & $|s_{0}\left\rangle \right.$  & $|s_{+1}\left\rangle \right.$  & Alice  & Bob  & $|s_{0}\left\rangle \right.$  & $|s_{+1}\left\rangle \right.$\tabularnewline
\hline 
\hline 
H  & H  & $\uparrow$  & $\uparrow$  & $+$  & $+$  & $|n_{+}\left\rangle \right.$  & $\times$\tabularnewline
\hline 
V  & V  & $\uparrow$  & $\uparrow$  & $-$  & $-$  & |$n_{-}\left\rangle \right.$  & $\times$\tabularnewline
\hline 
H  & V  & $\downarrow$  & $\downarrow$  & $+$  & $-$  & $\times$  & $|n_{+}\left\rangle \right.$\tabularnewline
\hline 
V  & H  & $\downarrow$  & $\downarrow$  & $-$  & $+$  & $\times$  & |$n_{-}\left\rangle \right.$\tabularnewline
\hline 
\end{tabular}
\par\end{centering}
\caption{\label{tab:outcomes}Possible states of the electron and nuclear spin
of the protocol of Fig. \ref{fig:Single_memory protocol} right before
the measurement. Here $|n_{\pm}\left\rangle \right.=\frac{1}{\sqrt{2}}\left(|\uparrow\left\rangle \right.\pm|\downarrow\left\rangle \right.\right)$
and the cross symbol $\text{\ensuremath{\times}}$ stands for an impossible
event in the ideal case. }
\end{table}

 \bibliographystyle{apsrev4-1}
\bibliography{bib1}

\end{document}